# Curvature-mediated spin textures in magnetic multi-layered nanotubes

Elisabeth Josten[1*,§], David Raftrey[2,3,§], Aurelio Hierro-Rodriguez[4], Andrea Sorrentino[5], Lucia Aballe[5], Marta Lipińska-Chwałek[1,6], Thomas Jansen[7], Katja Höflich[8,a], Hanno Kröncke[8], Catherine Dubourdieu[8,9], Daniel E. Bürgler[7], Joachim Mayer[1,6], Peter Fischer[2,3*]

[1]*Ernst Ruska-Centre for Microscopy and Spectroscopy with Electrons (ER-C), Forschungszentrum Jülich GmbH, 52425 Jülich, Germany*

[2]*Materials Sciences Division, Lawrence Berkeley National Laboratory, Berkeley, CA 94720 USA*

[3]*Department of Physics, University of California Santa Cruz, Santa Cruz, CA 95604 USA*

[4]*Departamento de Física, Universidad de Oviedo, 33007 Oviedo, Spain*

[5]*ALBA Synchrotron Light Facility, 08290 Cerdanyola del Vallès, Spain*

[6]*Central Facility for Electron Microscopy, RWTH Aachen University, 52074 Aachen, Germany*

[7]*Peter Grünberg Institute 6 and Jülich Aachen Research Alliance, Forschungszentrum Jülich GmbH, 52425 Jülich, Germany*

[8]*Helmholtz-Zentrum Berlin für Materialien und Energie, 14109 Berlin Germany*

[9]*Freie Universität Berlin, Physical Chemistry, 14195 Berlin, Germany*

[a]*present address: Ferdinand-Braun-Institut GmbH, Leibniz-Institut für Höchstfrequenztechnik, 12489 Berlin, Germany*

[§]*These authors contributed equally to this work.*

*Corresponding authors: e.josten@posteo.de, PJFischer@lbl.gov*

**Abstract**

The scientific and technological exploration of artificially designed three-dimensional magnetic nanostructures opens the path to exciting novel physical phenomena, originating from the increased complexity in spin textures, topology, and frustration in three dimensions. Theory predicts that the equilibrium magnetic ground state of two-dimensional systems which reflects the competition between symmetric (Heisenberg) and antisymmetric (Dzyaloshinskii-Moriya interaction (DMI)) exchange interaction is significantly modified on curved surfaces when the radius of local curvature becomes comparable to fundamental magnetic length scales.

Here, we present an experimental study of the spin texture in an 8 nm thin magnetic multilayer with growth-induced in-plane anisotropy and DMI deposited onto the curved surface of a 1.8 µm long non-magnetic carbon nanowire with a 67 nm radius. Using magnetic soft x-ray tomography the three-dimensional spin configuration in this nanotube was retrieved with about 30nm spatial resolution. The transition between two vortex configurations on the two ends of the nanotube with opposite circulation occurs through a domain wall that is aligned at an inclined angle relative to the wire axis. Three-dimensional micromagnetic simulations support the experimental observations and represent a visualization of the curvature-mediated DMI. They also allow a quantitative estimate of the DMI value for the magnetic multilayered nanotube.

**Main**

The properties, behavior and functionalities of magnetic materials are determined by their microscopic domain structure, which resembles the energetic ground state, and results from the competition of fundamental interactions, such as the symmetric Heisenberg exchange, which favors parallel or antiparallel alignment between neighboring spins, and the antisymmetric exchange, i.e., the Dzyaloshinskii-Moriya interaction (DMI) that can lead to the formation of non-collinear spin textures. Chirality, which is a general feature in many natural phenomena, and topology are signature characteristics of non-collinear spin textures such as vortices [1], skyrmions [2, 3], or Hopfions [4-6], and have thus become a major focus of research in nanomagnetism. Whereas understanding the effect of lower dimensionalities via confinement or proximity in magnetic materials led e.g., to studies and applications of primarily 1D or 2D systems, such as thin magnetic films, multilayers or patterned nanostructures, the extension into the third dimension, while still controlling the building blocks at the nanoscale, allows to create novel spin textures with interesting static and dynamic properties [7, 8]. Recent examples include studies of skyrmion tubes and strings [9-11], the dynamics of vortex cores along the thickness of a soft magnetic nanodisk [12], the formation of chiral bobbers [13, 14], the stabilization of Hopfions [6, 15], or the creation of an artificial double-helix spin textures in bespoke magnetic nanowires [16].

The extension into 3D includes planar 2D magnetic systems that reside on non-flat surfaces. Here, the interplay between topology, magnetic ordering, and geometry leads to additional effects that do not exist on flat surfaces. The curvature of the surface becomes an important design parameter, as e.g., the magnetic ground state of a simple Heisenberg magnet will be modified and depends on the curvature [17]. The local radius, particularly if its magnitude becomes comparable to fundamental magnetic length scales, such as exchange lengths, induces effective magnetic interactions, such as magnetic anisotropy or DMI [18]. As shown in Fig. 1a, the ground state of neighboring spins interacting through exchange interaction would no longer be the parallel alignment, but would become non-collinear, which on a flat surface indicates an antisymmetric exchange interaction, i.e., a DMI. In case of an axially symmetric surface, such as the surface of an elongated straight nanowire, the alignment of two neighboring spins would depend on whether their direction was along the nanowire axis or tangential to the nanowire radius, as in the former case they would be collinear and non-collinear in the latter case (Fig. 1a).

In this work, we report an experimental study of a bespoke magnetic multilayer system with intrinsic DMI, that was deposited onto the surface of a non-magnetic nanowire, resulting in a multi-layered nanotube that allowed to explore the combination of intrinsic and curvature induced effects on the resulting spin texture. To characterize the full three-dimensional magnetic configuration, we performed magnetic soft x-ray tomography [19-21]. We observe vortex textures with opposite chirality on both ends of the nanotube reminiscent of spin textures in thin nano-rings. As predicted from recent theory [18], we find a domain wall which is inclined with respect to the axis of the nanotube thus facilitating the transition between the different vortex chiralities. Micromagnetic simulations that include DMI confirm both the observed curvature induced spin textures, and support a quantitative estimate of effective DMI values from the measured inclination angle.

**Experimental details**

Using focused electron-beam-induced deposition (FEBID) [22-26] as a direct writing process, a 1.8 μm long and 67 nm radius non-magnetic carbon nanowire was grown on a silicon wafer and encapsulated by a 9 nm aluminum oxide ($Al_2O_3$) passivation layer using atomic layer deposition (ALD) to avoid thermal degradation of the nanowire. After depositing a 1 nm MgO buffer layer, a magnetic multilayer system consisting of five repeats of a <Pt (0.35 nm)/ Co (0.35 nm)/ Ni (0.25 nm)> trilayer system was deposited by magnetron sputtering onto the surface of the C nanowire, and capped with another 2 nm MgO layer (Fig. 1b). Since the sputter beam was oriented at an angle of about 30° to the wire axis, this led to a growth-induced in-plane anisotropy [27], which impacted the stabilization of the observed spin textures. Cross-sectional STEM images (Fig. 1c) confirmed the circumferential uniformity of the multilayer on the curved substrate.

Magnetic soft x-ray tomography with about 30nm spatial resolution using the full-field Transmission soft X-ray Microscopy (TXM) endstation MISTRAL [28] at ALBA in Barcelona/Spain was used to image the 3D spin texture of the 1.8 µm long magnetic multi-layered nanotube, which was affixed to a 50 nm thin $Si_3N_4$ membrane, to allow for sufficient x-ray transparency (Fig. 1c). X-ray magnetic circular dichroism (XMCD) provided magnetic contrast at the Co $L_3$ and $L_2$ edges at 778.6 eV and 793.8 eV, resp. The tomographic data set consisted of an angular series of images of the magnetic domain structure in the multi-layered nanotube recorded at remanence and under ambient conditions and was obtained by rotating around the wire axis in the range from -64° to 64°. Since the propagation direction of the incoming circularly polarized x-ray beam was orthogonal to the rotation axis, the domain structure in each individual image shows the magnetization component that is orthogonal to the axis of the nanowire. Fig. 2a shows the schematics of the experiment and Fig. 2b two representative magnetic images at the $L_3$ and $L_2$ edges at a single angle. The difference between those two images, represented by $\log\left(\frac{L_3}{L_2}\right)$ (Fig. 2b right panel) shows already a reversed contrast on the left and right edge of the nanotube indicative of a domain structure that points into opposite directions at the upper and lower part of the nanotube.

The full 3D spin texture (Fig. 2c) was retrieved by applying a dedicated reconstruction algorithm [19] (for details see Materials and Methods). Towards the top and bottom ends of the nanotube two domain structures are visible which circulate in a transversal plane to the wire (=rotation) axis with opposite sense of circulation. Fig. 2d and 2f displays two slices from the reconstructed data in those areas of the nanotube. The opposite chirality, i.e. clockwise and counterclockwise circulation is reminiscent of vortex patterns found in magnetic nanorings [29]. Vortex states have also been observed in hexagonal permalloy nanotubes [27]. Interestingly, due to the opposite chirality on both ends, the nanotube is forced to reverse its chirality along the nanotube by forming a domain wall. The experimental 3D data allow detailed insight into the spin textures of this transition region, e.g., the transition width is found to be about 65 nm (see Fig 3d). An analogous slice for this region, which is shown in Fig. 2e resembles the spin texture of an onion state, which has been discussed in permalloy nanorings [29]. The most striking observation is that the domain wall is aligned at an inclined angle of about 70° relative to the wire axis. Micromagnetic simulations using the OOMMF package [30] with interfacial DMI show that such an angle would not be expected for a flat surface, and in particular not for a nanotube without intrinsic interfacial DMI interacting multilayers (Fig. 3e). Therefore, the experimental observation of the inclination angle of the domain wall in the multilayered nanotube is strong evidence of the impact of the surface curvature on the intrinsic interfacial DMI as recently predicted from theory [18].

**Discussion**

To obtain a deeper understanding of the observed spin textures and in particular the impact of curvature, micromagnetic simulations using OOMMF were performed over a range of geometric and material parameters. Those indicate that the ground state of the system is strongly influenced by the geometrical aspect ratio of the tube as well as the material parameters, e.g., anisotropy. Whereas for tubes with a small radius the ground state is a uniform magnetization aligned with the axis of the tube, for larger radii, a vortex state is favored. Further, in addition to the radial dependence of the ground state, an additional tangential anisotropy, which was added to model the growth induced anisotropy, favors the vortex state. The character of the domain wall between the two vortices is modified by the DMI, and therefore provides quantitative insight into the interplay of chirality and curvature.

**Effect of Curvature on DMI**

Generally, in flat thin films interfacial DMI does not affect in-plane spin textures [31] as there is no symmetry breaking in the plane and therefore no preferred in-plane spin orientation. However, in a nanotube, the DMI is able to influence the in-plane vortex state because the surface is curved (Fig. 1a). In the case of interfacial DMI, the free energy resulting from the DMI is described by

$$E_{DMI} = -\vec{D_{ij}} \cdot (\vec{S_i} \times \vec{S_j})$$

where the DMI vector $\vec{D_{ij}}$ always lies in-plane. This is enforced by $\vec{D_{ij}} = d(\vec{U_{ij}} \times \hat{n})$ where $\vec{U_{ij}}$ is the displacement vector from one spin site to another, and $\hat{n}$ is the unit normal vector. The cross product of two in-plane spins points orthogonal to the plane, with no component in the direction of $\vec{D_{ij}}$. This means that the DMI will not contribute to the magnetic free energy. However, in the case of a cylindrical system in a vortex state, the curvature induced change in the direction of neighboring tangential spins will give their cross product an in-plane component, which therefore leads to a non-zero contribution from the DMI.

**Vortex structures at the top and bottom**

In an axially symmetric nanotube there are two possible ground states, the axial state and the vortex state. Results from micromagnetic simulations, which are included in the Supplementary Materials show that the axial state is preferred for smaller radii, but since at both ends there will be magnetic charges, which increase with increasing radius, the vortex state with uniform chirality is energetically preferred for larger radii. With decreasing length of the nanotube, the vortex state is preferred as well, which is consistent with the vortex state found in thin nanorings [29]. As mentioned above, in our system, the tangential anisotropy induced by the sputtering process at an oblique angle also promotes the formation of the vortex state. The formation of two vortices with opposite chiralities (Fig. 2c), which is observed here is unexpected, but can be assigned to a field-dependent demagnetization procedure that was applied to our system prior to the tomographic X-ray microscopy measurements. Most interestingly, this requires the formation of a domain wall to facilitate the transition between the two vortices.

**Domain wall in-between vortices**

To explain the observed inclination angle of the domain wall, one needs to consider the impact of the DMI (Fig. 4a). Simulations without DMI show that the transition between the two vortex chiralities occurs via a domain wall that is in the axial state and is confined to a plane perpendicular to the nanotube axis (Figs. 3e, 4c). In contrast, our experimental observation shows a transition region of about 65 nm in width, where the domain wall plane is no longer perpendicular to the nanotube axis (Fig. 4b). Simulations suggest that the onion-type transition which results from the inclination angle of the domain wall is a signature of the DMI on a curved background. Fig. 3f and g show digitally unwrapped images of the nanotube for the domain wall transition region from simulation with no DMI (Fig. 3e), with DMI (Fig. 3f), and experiment (Fig. 3g). Since the exchange length of our multilayer system is around 7 nm, the curvature, i.e. the inner radius of the nanotube (67 nm) is sufficiently small to affect the spin texture [18]. This is in agreement with simulations where the ratio $\frac{l_{exc}}{r}$ between magnetic exchange length $l_{exc}$ and curvature radius $r$ is varied and which confirm that the vortex state in a nanotube is favored for $\frac{l_{exc}}{r} = 0.1$. For details see Fig S1c in the Supplementary Information.

The character of the domain wall can be described as an in-plane Néel wall since, as theoretically supported, both the intrinsic interfacial DMI and the curvature induced effects promote a Néel type domain wall [18].

The novel feature arising from the combination of intrinsic interfacial DMI and curvature is the orientation of the domain wall which was measured to be an angle of 70° relative to the wire axis. Since our micromagnetic simulations suggest that with increasing DMI strength, the inclination angle increases as well, this allows an estimate on the value of the curvature mediated DMI value (Fig 4a). Our experimental data are consistent with a DMI value of $1 mJ/m^2$, which is comparable to DMI values found in flat films [32-35], where the interface between a magnetic material with a heavy element induces the DMI. Here, it means that the value of the DMI constant is not modified, but there is an additional curvature-induced component to the DMI vector due to the non-flat geometry.

**Conclusions**

Through magnetic soft X-ray tomography with nanoscale spatial resolution we have determined the curvature induced spin texture in a magnetic multilayer system that was shaped into a nanotube by being deposited onto the surface of a non-magnetic nanowire. We have found vortex structures with opposite chirality on both ends of the tube, and a domain wall enabling the transition between those vortices. The domain wall is contained in a plane whose normal forms an angle with the nanotube axis. This behavior results from the combination of the intrinsic interfacial DMI present in the magnetic heterostructure, and the curved geometry of the nanotube. The value of the inclination angle is a quantitative measure of the interfacial DMI strength, and indicates a curvature-mediated impact of the DMI.

The effect of curvature is a new parameter that can be exploited in order to design future technological applications of nanoscale devices with new functionalities, and should also be taken into account in non-perfect magnetic 1D and 2D systems, e.g., where local structural defects or imperfect interfaces create locally curved geometries. Our work provides evidence that advanced synthesis, modelling and characterization tools have reached a level of maturity to enter this path.

**Methods**

**Sample synthesis and structural characterization**

To synthesize a curved magnetic multilayer, specifically a multi-layered magnetic nanotube with axial symmetry, a regular multilayer was deposited onto an axial symmetric nanowire. The substrate was a non-magnetic carbon nanowire with 67 nm radius and 1.8 µm length, which was fabricated in a direct writing process using focused electron-beam-induced deposition (FEBID) [22-24, 26] on a silicon wafer using a Zeiss Crossbeam 340 KMAT microscope with phenanthrene as carbon precursor. Typical deposition parameters employ an acceleration voltage of 15 kV and electron beam currents in the range of 50 – 500 pA. To avoid thermal degradation during the subsequent sputtering process, the carbon core was passivated by a 9 nm aluminum oxide ($Al_2O_3$) shell performing a thermal atomic layer deposition (ALD) process in an Oxford FlexAL system. The ALD process was optimized to run at a lower deposition temperature of 120° C. The used precursors were trimethylaluminum (TMA) and water at a pressure of 80 mTorr with 150 sccm argon purging in-between.

After depositing a 1nm MgO seed layer, a magnetic multilayer system consisting of five repetitions of a <Pt (0.35 nm)/ Co (0.35 nm)/ Ni (0.25 nm)> trilayer was deposited via magnetron sputtering, and capped with another 2 nm MgO protective layer.

During sputtering deposition, the Ar pressure was between 7 – 8 x $10^{-3}$ mbar. To ensure uniform deposition on the outer surface of the nanowires, all layers were deposited at an oblique angle of about 30° to the wire axes. Circumferential homogeneity of the layer thicknesses was achieved by low sputtering rates (0.01 - 0.06 nm/s) and rotation of the sample at a constant speed of 8 rpm around the wire axis.

Individual nanotubes were extracted from the silicon substrate on which the nanotubes were fabricated using a dual beam system FEI Helios NanoLab 400S workstation equipped with a scanning electron microscope (SEM) and a focused ion beam (FIB). A rotating micromanipulator was used for transport, which allows for a precise placement of the tube on the silicon nitride membrane. The cutting of the nanotube from the substrate yielded precise edges of the tubes. Finally, the extracted sample was fixed with carbon on the membrane and the result was proven by scanning electron microscopy (SEM), which was carried out at 10 kV, 0.1 nA and under an angle of 45° (Fig.1 c).

For structural characterization a cross-sectional lamella of the nanotube was prepared using also the dual-beam FEI Helios Nanolab 400S. Scanning Transmission Electron Microscopy (STEM) was carried out at 200 kV on an FEI Titan G2 TEM equipped with a Schottky field emission gun and a CEOS DCOR Cs probe corrector (Ernst Ruska-Centre for Microscopy and Spectroscopy with Electrons (ER-C) et al. (2016). FEI Titan G2 80-200 CREWLEY. *Journal of large-scale research facilities*, 2, A43).

**Magnetic soft x-ray tomography**

To image the spin texture in the nanotube, magnetic transmission soft x-ray microscopy (MTXM) using the MISTRAL endstation at ALBA in Barcelona/Spain was used, which records full field images using X-ray Magnetic Circular Dichroism (XMCD) as magnetic contrast. An MTXM system is built in analogy to an optical microscopy consisting of the source, a condenser optics, a high-resolution objective lens, which for x-rays is a Fresnel zone plate (FZP), and a two-dimensional back-illuminated CCD detector. At the MISTRAL beamline [28], where the MTXM images were recorded, the optical system consists of a plane grating monochromator (PGM) which provides monochromatic x-rays. A capillary condenser system collimates the photon beam and provides a uniform illumination of the specimen. An image of the transmitted photons is formed by the downstream high-resolution FZP of 25 nm outermost zone width and directly recorded by a CCD device. The spatial resolution in an MTXM is primarily determined by the quality of the imaging FZP and has shown to be able to reach down to less than 10 nm [36]. The FZP at MISTRAL provides about 20-30 nm spatial resolution with a half-pitch of 26nm [37]. MTXM detects directly the transmitted photons and therefore, if the incident photon intensity is known, the actual x-ray absorption coefficient can be derived following Beer's Law. The XMCD contrast in MTXM detects the magnetization component that is projected onto the photon propagation direction. In general, MTXM can image magnetic domains in varying external magnetic fields, however, for this study all images were recorded in remanence.

The concept of magnetic soft x-ray tomography is to record an angular series of MTXM images, which are then used to reconstruct the 3D spin texture from those projections. Since the accessible angular range at MISTRAL is limited by the geometry of the microscope, data were recorded only over a span of 124° (-64° to 64°), with the axis of rotation aligned parallel to the axis of the nanotube. The limited angular range lead to a loss of information and consequently spatial resolution in parts of the sample that would be captured by a full 180° tilt series.

A reconstruction of all three components of the magnetization from an angular scan around a single axis is not possible, as only the magnetization components that are in the plane perpendicular to the rotation axis can be accessed. However, due to the particular symmetry of the studied structure, its magnetic configuration, and the selected tomographic setup, where rotation and nanotube axes are parallel, the chirality and propagation direction of the x-rays is parallel to the vortex magnetization, and therefore the main components of the magnetization within the vortex states can be reconstructed already with a single angular scan.

The spatial resolution for magnetic soft x-ray tomography was around 30 nm, which is slightly larger than the resolution in the individual images, but still sufficient to confirm the vortex state around the circumference of the tube, and to resolve the features of the spin texture that are influenced by curvature. Magnetic contrast at each angle in the tilt series was determined by taking the difference between the intensity $I$ at the Co $L_3$ and $L_2$ edges, $I = \log\left(\frac{L_3}{L_2}\right)$. A Gaussian filter was applied to the raw magnetic contrast data in order to reduce noise.

The three dimensional rendering was produced from the two dimensional magnetic contrast files using an algebraic reconstruction algorithm [19]. The algorithm calculates the volume magnetization by solving a system of linear equations of the form $y^\phi - A^\phi x = 0$ for each component of the magnetization. Here the components of the three-dimensional magnetization reconstruction, $x$, is projected into the two-dimensional data $y^\phi$, by a projection matrix $A^\phi$, where $\phi$ is the projection angle. Each pixel on each projection image provides one condition for the system of equations. The magnetization is then reconstructed by iteratively solving the system of equations.

**Micromagnetic simulations**

The micromagnetic simulations were performed using the OOMMF micromagnetic simulation package [30]. Dipole-dipole interaction was included in the simulation, and the interfacial DMI module was modified to accommodate the curved tube geometry. The cell size was chosen as 4 nm x 4 nm x 4 nm. The material parameters used for the DMI D, exchange stiffness A, in-plane anisotropy K, and saturation magnetization $M_S$ were A = $10^{-11}$ J/m, K = $2 \cdot 10^4$ J/m³, $M_S$ = $700 \cdot 10^3$ A/m, and D was varied between $0.2 \cdot 10^{-3}$ and $1.4 \cdot 10^{-3}$ J/m², respectively.

**Acknowledgements**

We like to thank Maximilian Kruth for help with transferring the nanotubes using the FIB, and the CoreLab Correlative Microscopy and Spectroscopy (CCMS) at the Helmholtz-Zentrum Berlin for supporting the direct electron beam writing of the carbon nanowires.

This work was funded by the U.S. Department of Energy, Office of Science, Office of Basic Energy Sciences, Materials Sciences and Engineering Division under Contract No. DE-AC02-05-CH11231 (Non-equilibrium magnetic materials program MSMAG).

This research includes experiments that were performed at MISTRAL beamline at ALBA Synchrotron in collaboration with ALBA staff.

AH-R. acknowledges the support from European Union's Horizon 2020 research and innovation program under Marie Skłodowska-Curie grant ref. H2020-MSCA-IF-2016-746958 and from the Spanish AEI under project ref: PID2019–104604RB/AEI/10.13039/501100011033.

**Author information**

**A) Affiliations**

**Ernst Ruska-Centre for Microscopy and Spectroscopy with Electrons (ER-C), Forschungszentrum Jülich GmbH, 52425 Jülich, Germany**

Elisabeth Josten, Marta Lipińska-Chwałek, Joachim Mayer

**Materials Sciences Division, Lawrence Berkeley National Laboratory, Berkeley, CA 94720 USA and**

**Department of Physics, University of California Santa Cruz, Santa Cruz, CA 95604 USA**

David Raftrey, Peter Fischer

**Departamento de Física, Universidad de Oviedo, 33007 Oviedo, Spain**

Aurelio Hierro-Rodriguez

**ALBA Synchrotron Light Facility, 08290 Cerdanyola del Vallès, Spain**

Andrea Sorrentino, Lucia Aballe

**Central Facility for Electron Microscopy, RWTH Aachen University, 52074 Aachen, Germany**

Marta Lipińska-Chwałek, Joachim Mayer

**Peter Grünberg Institute 6 and Jülich Aachen Research Alliance, Forschungszentrum Jülich GmbH, 52425 Jülich, Germany**

Thomas Jansen, Daniel E. Bürgler

**Helmholtz-Zentrum Berlin für Materialien und Energie, 14109 Berlin Germany**

Katja Höflich, Hanno Kröncke, Catherine Dubourdieu

**Freie Universität Berlin, Physical Chemistry, 14195 Berlin, Germany**

Catherine Dubourdieu

**Ferdinand-Braun-Institut GmbH, Leibniz-Institut für Höchstfrequenztechnik, 12489 Berlin, Germany**

Katja Höflich

## B) Contributions

E.J. and P.F. designed and initiated the research, E.J., K.H., T.J., H.K., C.D., and D.B. discussed the design and preparation method of the sample, K.H., T.J., and H. K. synthesized the samples, M.L-C. and E.J. performed and analyzed the TEM measurements, L.A, A.S., D.R., A.H.-R., E.J. and P.F. performed the MTXM experiments at ALBA, D.R. and A.H.-R. analyzed the MTXM data, P.F., D.R. E.J., A.H.-R., L.A. discussed and wrote the manuscript. All authors commented on the final version of the manuscript.

## C) Corresponding authors

Elisabeth Josten (e.josten@posteo.de) and Peter Fischer (PJFischer@lbl.gov)

**Ethics declaration**

**Competing interests**

The authors declare no competing financial interests.

FIGURES

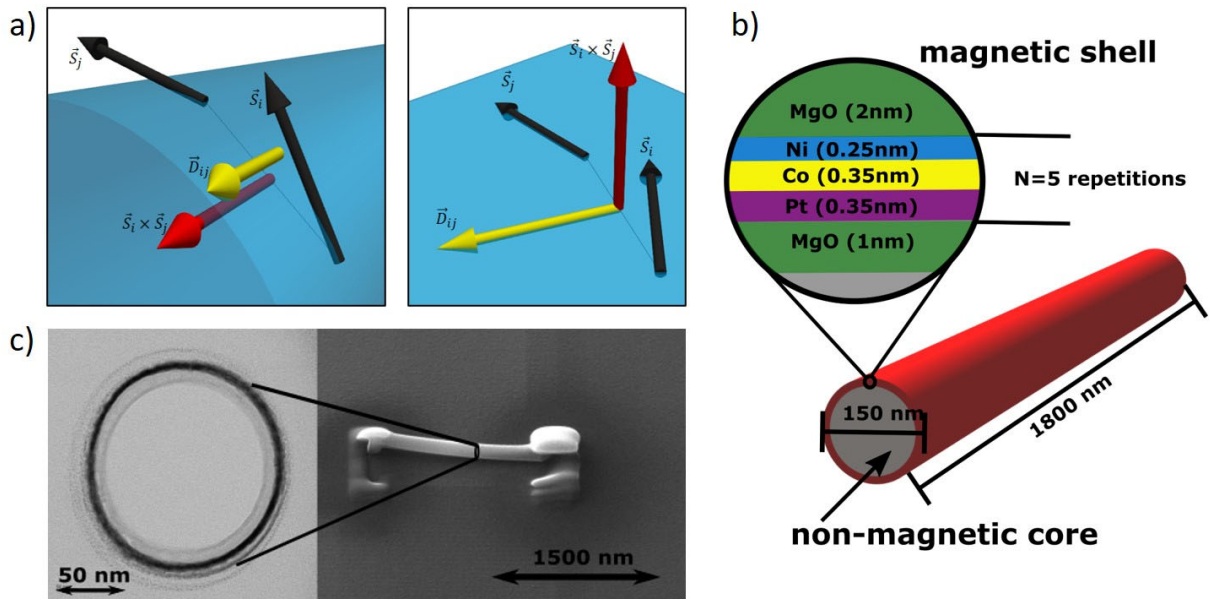

**Figure 1.** a) Effective DMI for neighboring spins on curved (left) and planar (right) surfaces. b) Schematic layout of the magnetic multi-layered nanotube. c) Cross-sectional bright-field STEM (left) and SEM (right) image of the deposited nanotube.

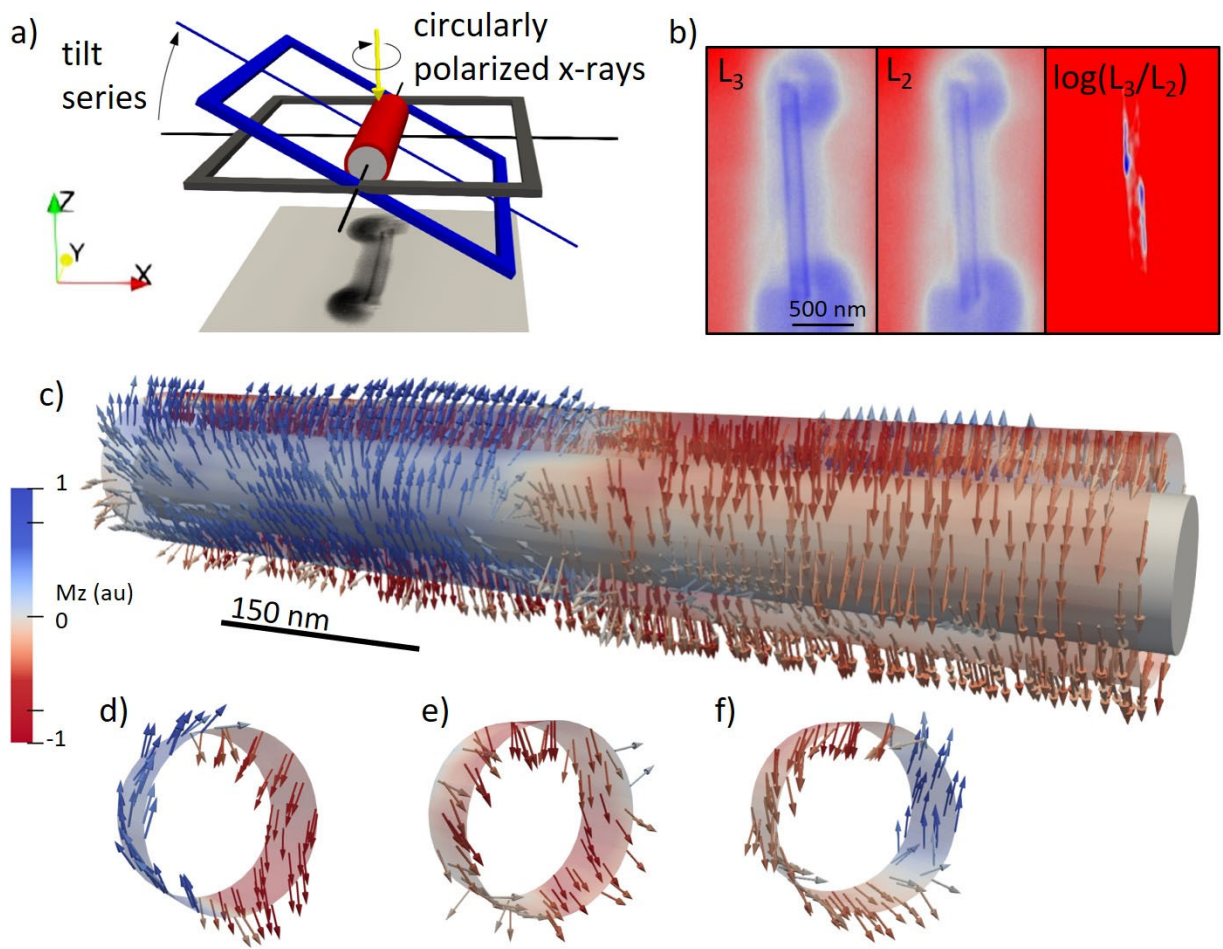

**Figure 2. Magnetic soft x-ray tomography** a) Schematics of the experimental setup, b) Raw experimental data recorded at the $L_3$, $L_2$ edges, and corresponding difference signal, c) Reconstructed 3D spin texture, d) - f) Slices from the reconstructed data taken at both ends d), f) and at the transition region e).

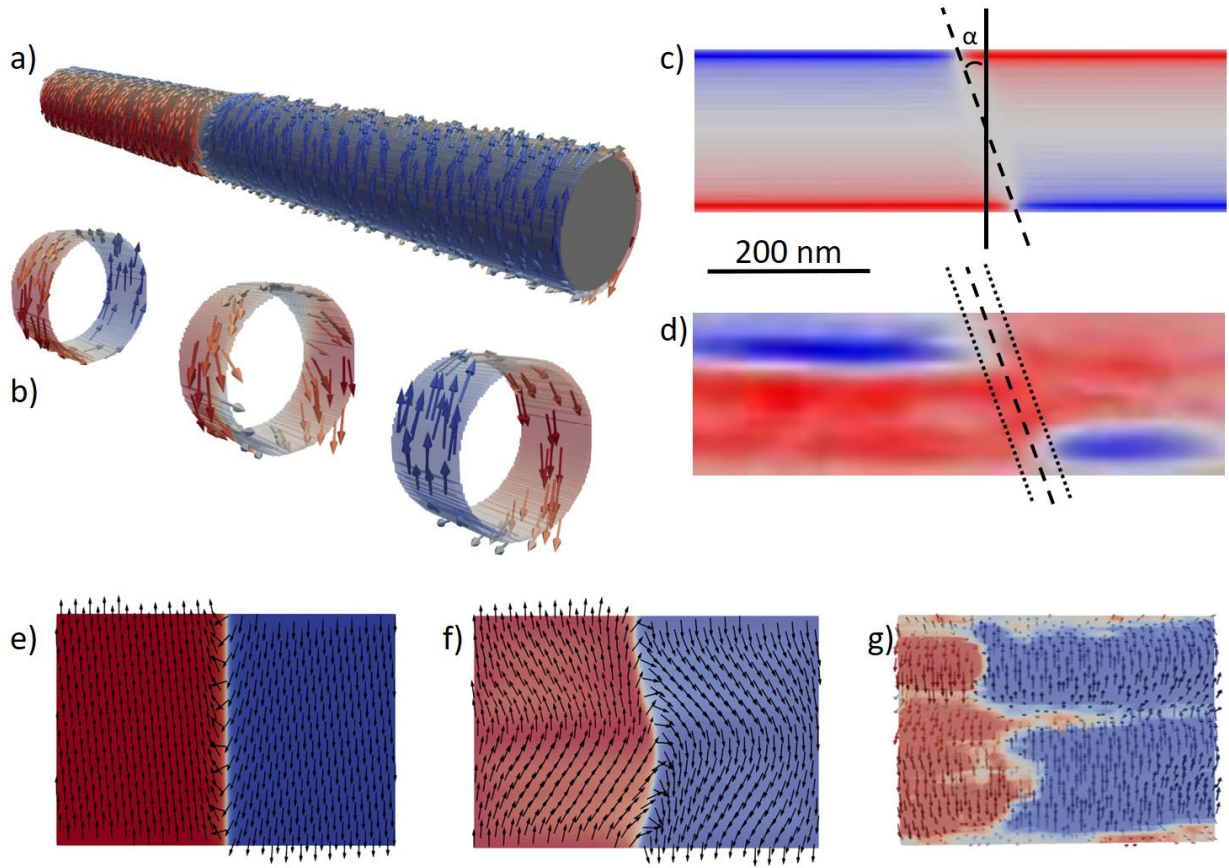

**Figure 3. Simulation and comparison with experiment:** a) Rendering of simulated nanotube magnetization. b) Slices showing the opposite chiralities on both ends and the transition region. c) Z-axis projection of simulated magnetization, d) Z-axis projection of experimental data. Dashed lines in c) and d) indicate the inclination of the domain wall (dotted line is experimental error). e), f) and g) Digitally unwrapped simulation of the nanotube e) without DMI, f) with DMI and g) experimental data.

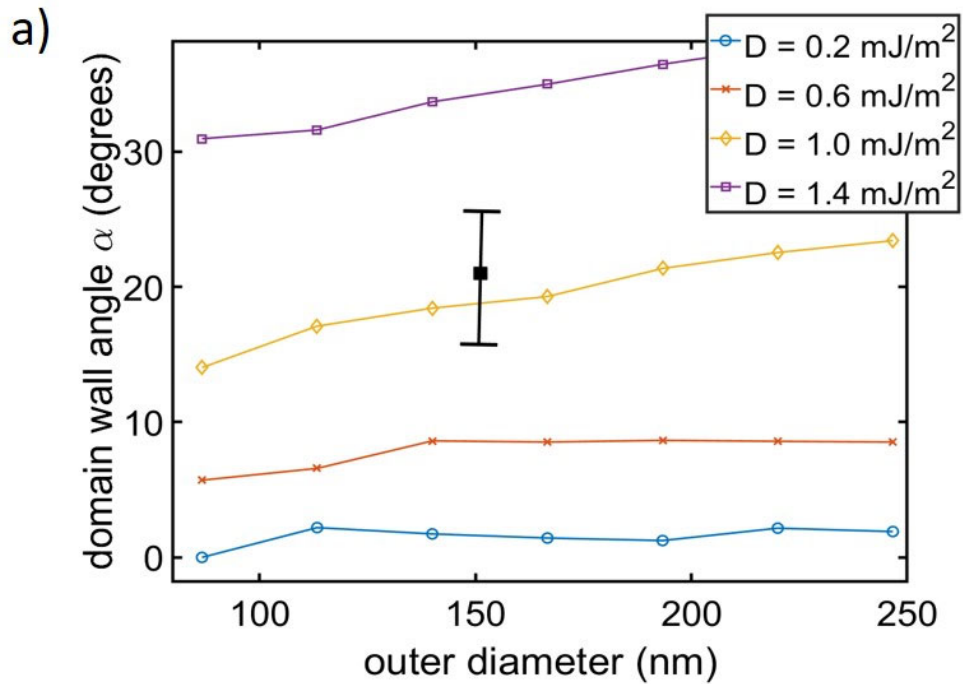

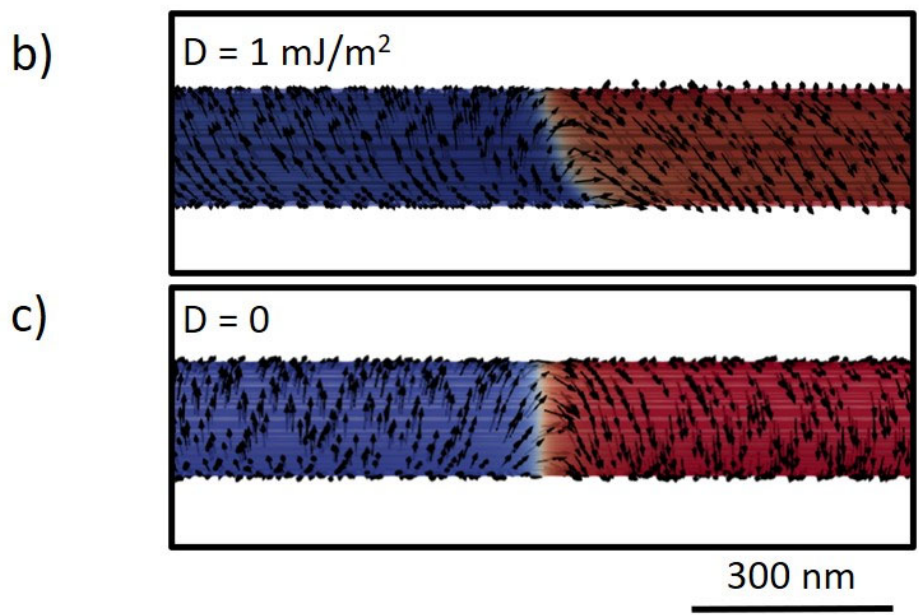

**Figure 4. Quantitative analysis of the DW angle.** a) Domain wall angle to nanowire axis as function of outer diameter for different DMI values. b) Simulated domain wall with DMI. c) Simulated domain wall without DMI.

# SUPPLEMENTARY INFORMATTION

# Curvature-mediated spin textures in magnetic multi-layered nanotubes

Elisabeth Josten[1*,§], David Raftrey[2,3,§], Aurelio Hierro-Rodriguez[4], Andrea Sorrentino[5], Lucia Aballe[5], Marta Lipińska-Chwałek[1,6], Thomas Jansen[7], Katja Höflich[8,a], Hanno Kröncke[8], Catherine Dubourdieu[8,9], Daniel E. Bürgler[7], Joachim Mayer[1,6], Peter Fischer[2,3]*

[1]*Ernst Ruska-Centre for Microscopy and Spectroscopy with Electrons (ER-C), Forschungszentrum Jülich GmbH, 52425 Jülich, Germany*

[2]*Materials Sciences Division, Lawrence Berkeley National Laboratory, Berkeley, CA 94720 USA*

[3]*Department of Physics, University of California Santa Cruz, Santa Cruz, CA 95604 USA*

[4]*Departamento de Física, Universidad de Oviedo, 33007 Oviedo, Spain*

[5]*ALBA Synchrotron Light Facility, 08290 Cerdanyola del Vallès, Spain*

[6]*Central Facility for Electron Microscopy, RWTH Aachen University, 52074 Aachen, Germany*

[7]*Peter Grünberg Institute 6 and Jülich Aachen Research Alliance, Forschungszentrum Jülich GmbH, 52425 Jülich, Germany*

[8]*Helmholtz-Zentrum Berlin für Materialien und Energie, 14109 Berlin Germany*

[9]*Freie Universität Berlin, Physical Chemistry, 14195 Berlin, Germany*

[a]*present address: Ferdinand-Braun-Institut GmbH, Leibniz-Institut für Höchstfrequenztechnik, 12489 Berlin, Germany*

[§]*These authors contributed equally to this work.*

*Corresponding authors: e.josten@posteo.de, PJFischer@lbl.gov*

- **Additional micromagnetic simulations**

a) axial state
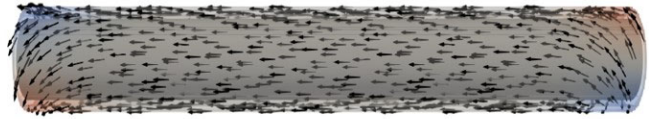

b) vortex state
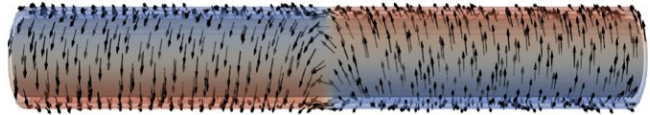

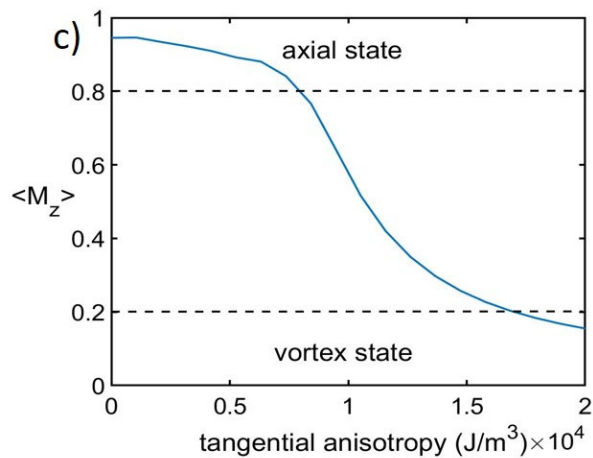
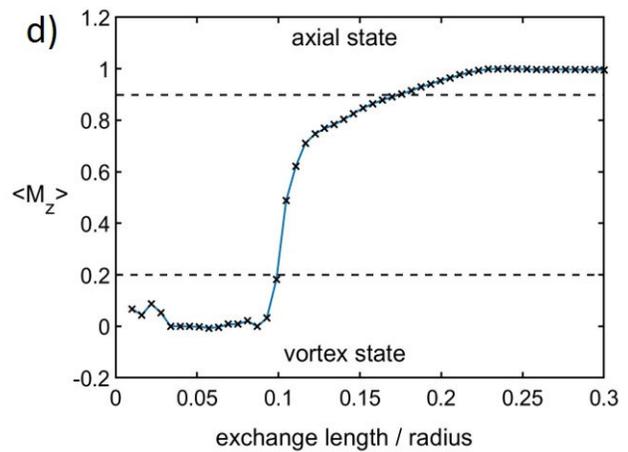

Micromagnetic simulation of the spin textures in the axial state and the vortex state are shown in Fig S1 a) and b). Quantitative simulation results are shown in Fig S1 c,d.

Fig S1 c) shows simulation results for the $M_z$ component of the magnetization as a function of the tangential anisotropy. They show that for a nanotube in our experimental geometry and for anisotropy values below *8 kJ/m³* the axial state is the ground state. In contrast, for a tangential anisotropy above *15 kJ/m³* the vortex state is favored.

The dependence on the ratio of magnetic exchange length/radius is shown in Fig S1 d). The axial state is favored for tubes with larger exchange length/radius aspect ratios, whereas the vortex state is the ground state for tubes with low exchange length/radius aspect ratios. The transition from a vortex to an axial ground state occurs when the ratio of exchange length to radius is greater than 0.1. For our experimental geometrical parameters we would expect an axial state without an additional tangential anisotropy.

- **Supplementary AVI movies**

  The supplementary movies illustrate the 3D results for the experiment and the relevant simulation.